\documentclass{aa}
\usepackage{graphicx}
\usepackage{natbib}
\usepackage{hyperref}
\usepackage{txfonts}
\begin{document}

   \title{BEAST begins: Sample characteristics and survey performance of the B-star Exoplanet Abundance Study
\thanks{Based on observations from the European Southern Observatory, Chile (Programmes 1101.C-0258 and 0103.C-0251).}
}


   \author{Markus Janson\inst{1} \and
          Vito Squicciarini\inst{2,3} \and
          Philippe Delorme\inst{4} \and
          Raffaele Gratton\inst{3} \and
          Micka{\"e}l Bonnefoy\inst{4} \and
          Sabine Reffert\inst{5} \and
          Eric E. Mamajek\inst{6,7} \and
          Simon C. Eriksson\inst{1} \and
          Arthur Vigan\inst{8} \and
          Maud Langlois\inst{9} \and
          Natalia Engler\inst{10}
          Ga{\"e}l Chauvin\inst{11,4} \and
          Silvano Desidera\inst{3} \and
          Lucio Mayer\inst{12} \and
          Gabriel-Dominique Marleau\inst{13,14,15} \and
          Alexander J. Bohn\inst{16} \and
          Matthias Samland\inst{1} \and
          Michael Meyer\inst{17}
          Valentina d’Orazi\inst{3} \and
          Thomas Henning\inst{15} \and
          Sascha Quanz\inst{10} \and
          Matthew Kenworthy\inst{16} \and
          Joseph C. Carson\inst{18}
          }

         \institute{Institutionen f{\"o}r astronomi, Stockholms Universitet, Stockholm, Sweden\\
              \email{markus.janson@astro.su.se}
        \and
           Dipartimento di Fisica e Astronomia ``Galileo Galilei'' Universit{\`a} di Padova, Padova, Italy
        \and
           INAF - Osservatorio Astronomico di Padova, Padova, Italy
         \and
           Univ. Grenoble Alpes, IPAG, Grenoble, France
         \and
           Landessternwarte, Zentrum f{\"u}r Astronomie der Universit{\"a}t Heidelberg, Heidelberg, Germany
         \and
           Jet Propulsion Laboratory, California Institute of Technology, Pasadena, CA, USA
         \and
           Department of Physics and Astronomy, University of Rochester, Rochester, NY, USA
        \and
            Aix Marseille Universit{\'e}, CNRS, LAM, Marseille, France
        \and
            CRAL, CNRS, Universit{\'e} Lyon, Saint Genis Laval, France
        \and
            ETH Zurich, Zurich, Switzerland
        \and
            Departemento de Astronom\'ia, Universidad de Chile, Santiago, Chile
        \and
            Center for Theoretical Astrophysics and Cosmology, Institute for Computational Science, University of Zurich, Zurich, Switzerland
        \and
            Institut f{\"u}r Astronomie und Astrophysik, Eberhard Karls Universit{\"a}t T{\"u}bingen, T{\"u}bingen, Germany
        \and
            Physikalisches Institut, Universit\"{a}t Bern, Bern, Switzerland
        \and
            Max Planck Institut f{\"u}r Astronomie, Heidelberg, Germany
        \and
            Leiden Observatory, Leiden University, Leiden, The Netherlands
        \and
            Department of Astronomy, University of Michigan, Ann Arbor, MI, USA
        \and
            College of Charleston, Charleston, SC, USA
             }

   \date{Received ---; accepted ---}

   \abstract{While the occurrence rate of wide giant planets appears to increase with stellar mass at least up through the A-type regime, B-type stars have not been systematically studied in large-scale surveys so far. It therefore remains unclear up to what stellar mass this occurrence trend continues. The B-star Exoplanet Abundance Study (BEAST) is a direct imaging survey with the extreme adaptive optics instrument SPHERE, targeting 85 B-type stars in the young Scorpius-Centaurus (Sco-Cen) region with the aim to detect giant planets at wide separations and constrain their occurrence rate and physical properties. The statistical outcome of the survey will help determine if and where an upper stellar mass limit for planet formation occurs. In this work, we describe the selection and characterization of the BEAST target sample. Particular emphasis is placed on the age of each system, which is a central parameter in interpreting direct imaging observations. We implement a novel scheme for age dating based on kinematic sub-structures within Sco-Cen, which complements and expands upon previous age determinations in the literature. We also present initial results from the first epoch observations, including the detections of ten stellar companions, of which six were previously unknown. All planetary candidates in the survey will need follow up in second epoch observations, which are part of the allocated observational programme and will be executed in the near future.}


\titlerunning{BEAST survey}
\authorrunning{M. Janson et al.}

   \maketitle
%

\section{Introduction}
\label{s:intro}

Studying planetary populations in a range of stellar environments is of critical importance for understanding their formation and early evolution. High-contrast imaging with adaptive optics (AO) is a valuable technique for studying the demographics of wide giant planets in this context, in virtually any kind of environment. Direct imaging also preferentially facilitates the study of young systems, representing an early and potentially pristine stage of the evolution of the system. Consequently, a large number of surveys have been performed to better understand this population and its distribution in a range environments such as around low-mass stars \citep[e.g.][]{delorme2012,bowler2015,lannier2016}, Sun-like stars \citep[e.g.][]{lafreniere2007,brandt2014,chauvin2015}, A-type stars \citep[e.g.][]{vigan2012,rameau2013a}, binaries \citep[e.g.][]{bonavita2016,asensio2018,hagelberg2020}, disc hosts \citep[e.g.][]{janson2013a,meshkat2017,lombart2020}, as well as broader surveys covering multiple such demographics \citep[e.g.][]{nielsen2019,desidera2020}. These efforts have resulted in the detections of several planets and low-mass substellar companions \citep[e.g.][]{marois2008,lagrange2010,kuzuhara2013,rameau2013b,macintosh2015,chauvin2017}, including some that appear to still be undergoing formation \citep{keppler2018,haffert2019,eriksson2020}. A clear trend that has emerged from such studies is that giant planets appear to be significantly more common around more massive stars throughout the M-type to A-type range \citep[e.g.][]{crepp2011,nielsen2019,vigan2020}. The extremes of this range are particularly informative in this context, so a critical question is how such relations extend to yet higher masses. Unfortunately, much less is known in the B-type stellar regime. Only one dedicated survey for planets around B-type stars has been performed \citep{janson2011}, which utilized a previous-generation AO system and featured a modest sample size.

In this paper, we outline the B-star Exoplanet Abundance Study (BEAST), which is a dedicated survey for planets around B-type stars with the SPHERE instrument \citep{beuzit2019} at the Very Large Telescope (VLT). The BEAST study is run as a so-called large programme, spanning several observing periods, and targets 85 stars in the Sco-Cen young stellar association. In parallel, we are also analysing archival data of individual B-stars in Sco-Cen that have been observed in other programmes. This has already led to the detection of the low-mass substellar circumbinary companion HIP 79098 (AB)b \citep{janson2019}. In this paper, however, we focus on the large programme that forms the core of BEAST.

While specific B-star surveys have been limited, individual late B-type objects have been observed in various programmes, which has already resulted in several detections of low-mass substellar companions \citep[e.g.][]{lafreniere2011,carson2013,cheetham2018}. This seems to imply that the trend of increasing wide giant planet frequency from M- to A-type stars continues at least up through $\sim$B9 types. Meanwhile, planet frequency as a function of stellar mass has also been examined in the context of radial velocity (RV) studies. While massive stars are difficult to study with RV on the main sequence (MS) because of their lack of narrow spectral lines, these stars expand and cool in the post-MS evolution resulting in more numerous and narrower lines, thereby making these star much more suitable for such studies. Radial velocity preferentially covers small orbital separations, and thus forms an excellent complement to direct imaging for probing planet frequency at a range of separations. Intriguingly, the RV studies show the same trend as direct imaging studies with an increasing giant planet frequency as a function of stellar mass, up to a mass of $\sim$2~$M_{\rm sun}$ \citep[][Wolthoff et al. in prep]{johnson2010,reffert2015}. However, in the mass range 2--3~$M_{\rm sun}$, the frequency turns over and starts decreasing with increasing stellar mass \citep{reffert2015}. Incidentally, a mass of 2.5~$M_{\rm sun}$ roughly corresponds to a spectral type of B9 on the MS, and thus marks the approximate transition between the A and B spectral type ranges.

The decrease in planet frequency in the 2--3~$M_{\rm sun}$ range in RV studies could be seen either as a formation-related or a migration-related issue. On one hand, discs around massive stars are probably more massive themselves, which probably benefits planet formation. Meanwhile, they experience higher doses of high-energy radiation from the central star than planets around lower-mass stars, so the disc also potentially dissipates faster. In relation to the observed apparent planet abundance turnover at $\sim$2~$M_{\rm sun}$, in RV studies \citep{reffert2015} this could be interpreted as the point at which the negative effects of the dissipation outweighs the positive effects of an enhanced mass reservoir. 

On the other hand, giant planets at sufficiently close separations to be observable with RV are generally expected to have formed at larger separations than their current orbit and subsequently migrated inwards in the disc \citep[e.g.][]{mordasini2009}. Theoretical work \citep{kennedy2008} predicts that the gradual evolution of the protoplanetary disc causes migration to slow down around more massive stars. Hence, the giant planet frequency may potentially keep increasing beyond a stellar mass of 2~$M_{\rm sun}$, but simultaneously, the planets may be increasingly prohibited from migrating into the inner parts of the system, where they could have been detectable with RV. Both of these possible scenarios make identical predictions for RV surveys. But they make diametrically opposite predictions for direct imaging, because the wide giant planet frequency would decrease in the 2--3~$M_{\rm sun}$ range if formation is halted, while the frequency would instead increase if migration is halted. A direct imaging survey such as BEAST is required to address this ambiguity.

Moreover, specific formation scenarios, such as disc instability, may occur preferentially in the outermost regions of massive discs encircling massive stars \citep{helled2014}. Accelerated disc dissipation would not be an issue in this case because disc instability occurs fast, likely in the first $10^5$ yr of the disc lifetime. Conversely, enhanced stellar irradiation would increase the temperature of the inner disc, possibly slowing down or even stopping migration \citep{rowther2020}, which can otherwise be very fast in unstable discs \citep{mueller2018}. As a result, a population of substellar companions formed by disc instability (gas giants as well as brown dwarfs) might  be completely absent in RV surveys, while it should be fully accessible to BEAST. The discovery of a coherent excess population of gas giants and brown dwarfs with wide orbits and a relatively top-heavy mass distribution would be a clear indication in favour of this distribution having been formed via disc instability. Early indications of two distinct population of directly imaged planets/brown dwarfs already exists around lower-mass stars \citep{nielsen2019,vigan2020} and may potentially be considerably more clearly distinguished in a survey of massive stars.

The BEAST survey is entirely focussed on the Sco-Cen region \citep{dezeeuw1999}, which is a young ($\sim$5--20 Myr) and relatively nearby ($\sim$120--150~pc) co-moving association of stars \citep{pecaut2016}, usually subdivided into the Upper Scorpius (USco), Upper Centaurus Lupus (UCL), and Lower Centaurus Crux (LCC) regions. There are several reasons for the choice of Sco-Cen as target region for the survey: Importantly, it offers one of the most favourable trade-offs between proximity and youth out of all known associations of stars \citep[e.g.][]{pecaut2016}. Both proximity and youth are important factors in facilitating detections with direct imaging, but they are partly mutually exclusive because younger stars are rarer and thus, on average, more distant. For example, there are more nearby young moving groups than Sco-Cen, but they are not as young, or not large enough to contain any significant number of B-type stars \citep[e.g.][]{zuckerman2004,gagne2018}. Conversely, there are younger associations and clusters than Sco-Cen, but they are more distant and often exhibit substantial levels of extinction. Sco-Cen is the most nearby coherent region that offers a large ($>$100) sample of B-type stars. 

By focussing on essentially all suitable stars in the Sco-Cen region, we acquire a sample that is as uniform as can be achieved in terms of ages and metallicities, which will benefit the statistical analysis of the full survey. Another crucial benefit to using the Sco-Cen region is that it is already heavily targeted for lower-mass (F and A-type) stars in the SHINE survey \citep{desidera2020}, as well as other smaller surveys using SPHERE. Hence, we will be able to statistically compare the BEAST sample with a lower-mass sample in the same region. This comparison effectively eliminates the impact of factors such as age and metallicity, since they are relatively uniform across the region and thus cleanly isolates stellar mass as the fundamental parameter to test planet occurrence rate against.

\section{BEAST sample}
\label{s:sample}

\subsection{Sample selection}
\label{s:selection}

Our initial master list for BEAST target selection consisted of all B-type stars identified as members of Sco-Cen with $>$50\% probability in a Bayesian kinematic membership study by \citet{rizzuto2011}. This input list contains 165 potential targets. We also added 11 targets that were identified in the \citet{dezeeuw1999} list of Sco-Cen objects, and for which the BANYAN \citep{gagne2014,gagne2018} code provided a $>$50\% probability of Sco-Cen membership. All individual targets were then further examined and removed if they fulfilled any of the following criteria: 

(1) While wide binaries and spectroscopic binaries were kept in the sample, physically bound binaries with intermediate separations of 0.1--6$^{\prime \prime}$ were removed. The reason for this choice is that secondaries in this separation range would have a detrimental effect on the contrast performance of SPHERE. Binaries in the relevant separation range are generally identified from the Washington Double Star \citep[WDS;][]{mason2001} survey. 

(2) Targets that have already been observed by SPHERE with comparable settings and comparable depth were also removed from the sample. This is done to avoid duplications in accordance with European Southern Observatories (ESO) policy. Our long-term plan is to eventually combine the BEAST survey with all other sufficiently deep SPHERE observations of B-type stars in Sco-Cen to yield a full census of the B-star wide giant planet population in this region. 

(3) Objects with a declination within approximately $\pm$3 deg from -24.6 deg (the latitude of Paranal observatory where SPHERE resides) were also removed. Targets in this declination band pass close enough to zenith (as seen from the telescope) that tracking becomes unfeasible near an hour angle of 0 h. This coincides with the time interval during which the target needs to be observed for angular differential imaging (ADI) purposes \citep{marois2006}. Hence, ADI cannot be performed with high efficiency for such targets. 

\begin{figure}[htb]
\centering
\includegraphics[width=8cm]{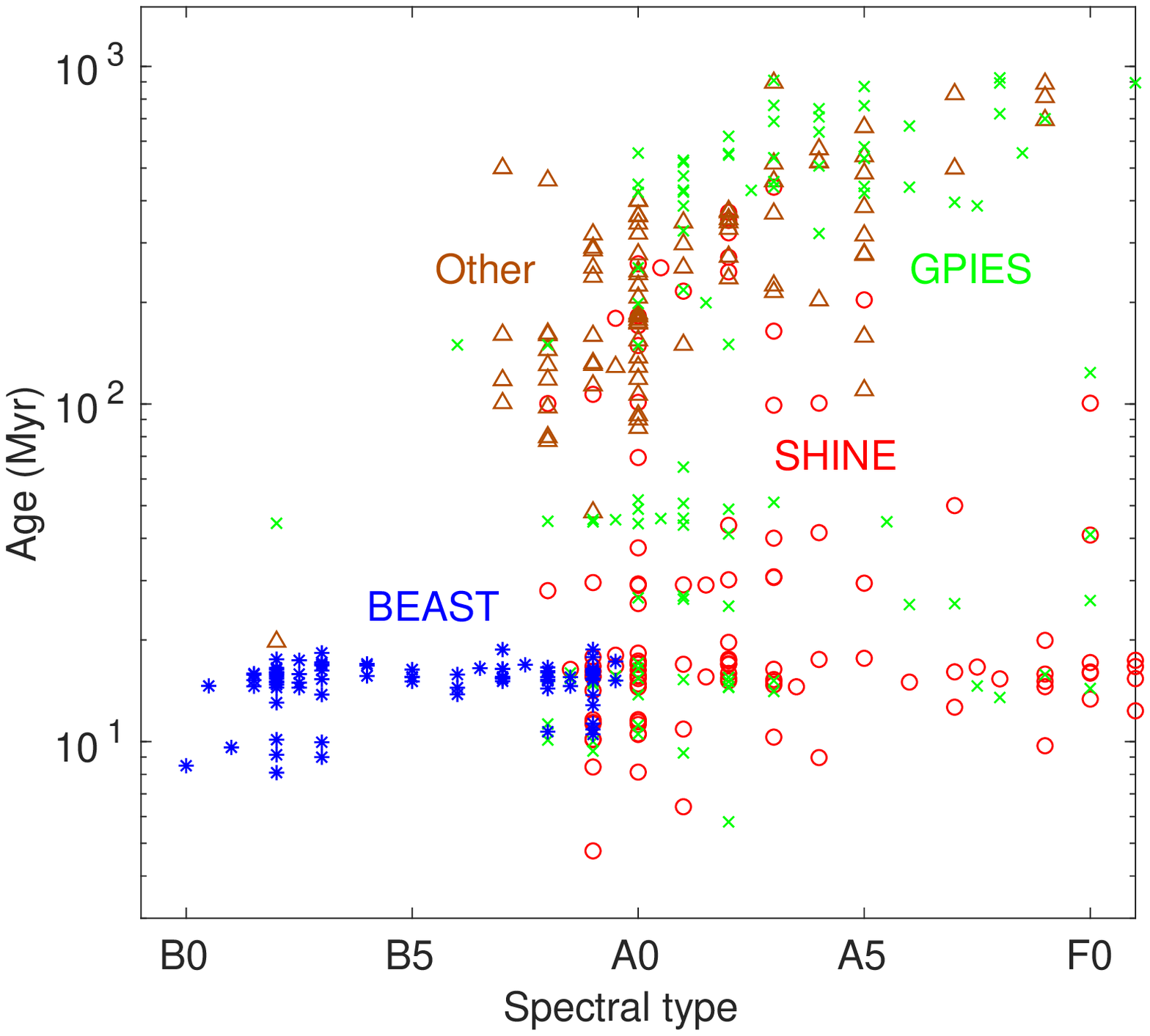}
\caption{Comparison of BEAST (blue asterisks), SHINE \citep{desidera2020}  (red circles), and GPIES \citep{nielsen2019} samples (green crosses). Also shown as brown triangles are targets from other previous surveys covering early-type stars \citep{janson2011,nielsen2013}. To enhance visibility, a Gaussian stochastic term with standard deviation of 1 Myr was added to all stellar ages in the figure. The massive stellar range covered in BEAST has not been previously systematically probed in direct imaging studies.}
\label{f:samples}
\end{figure}

Our final BEAST target sample consists of 85 B-type stars that fulfil all of the constraints for selection. The BEAST survey is the first to cover a large sample of young B-type stars -- a comparison to other direct imaging surveys in terms of spectral type and age is shown in Fig. \ref{f:samples}. In the following, we discuss the determination of the detailed characteristics for the targets in the sample.

\subsection{Membership of Sco-Cen}
\label{s:membership}

The majority of our sample originates from the \citet{rizzuto2011} Bayesian analysis of Sco-Cen membership, wherein the objects were determined as probable members of the association. However, with new astrometry from \textit{Gaia} DR2 \citep{brown2018} and more sophisticated Bayesian models \citep{gagne2018} that have become available since the onset of the survey, it is relevant to re-evaluate membership for each individual member. 

\begin{figure}[htb]
\centering
\includegraphics[width=8cm]{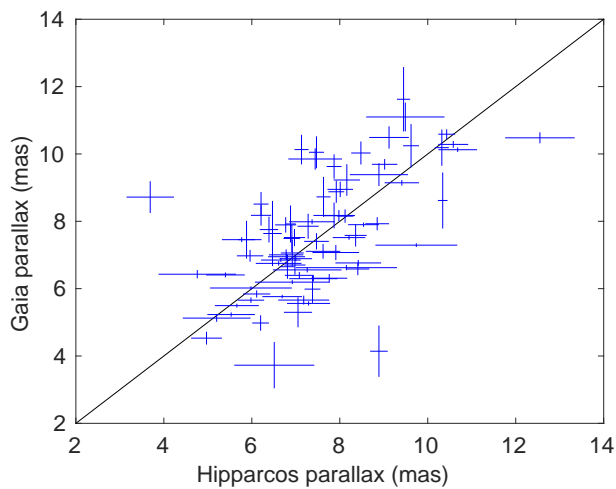}
\caption{Measured parallaxes from \textit{Hipparcos} vs. \textit{Gaia} for the targets in our sample. The solid diagonal line denotes a 1-to-1 relationship. The scatter is larger than captured by the estimated individual error bars, implying excess errors from, for example unresolved multiplicity and high target brightness.}
\label{f:plxplx}
\end{figure}

An important aspect in this analysis is the choice of astrometric catalogue. For the analysis in this paper, we consistently used \textit{Gaia} DR2. A comparison between \textit{Gaia} DR2 and the \citet{vanleeuwen2007} reduction of \textit{Hipparcos} data for the BEAST targets is shown in Fig. \ref{f:plxplx}. \textit{Gaia} has somewhat smaller formal error bars (median of 0.21 mas) than \textit{Hipparcos} (median of 0.34 mas). However, this does not necessarily translate to a higher accuracy, at least not in all cases, because the \textit{Gaia} DR2 reduction only attempts to fit the parallax and proper motion. Therefore any photocentre shifts due to binarity within the systems lead to systematic errors in the fitted quantities. The magnitude of the effect depends on the separation and flux ratio of the binary components. This is particularly relevant for BEAST, whose targets are early-type and thus have a relatively high multiplicity fraction \citep{duchene2013}. The target stars are also outside of the brightness range for which the DR2 reduction is optimized ($>$6 mag stars). We flagged many of the BEAST targets for astrometric excess errors, which may result from such effects. The scatter around a 1-to-1 relationship in Fig. \ref{f:plxplx} is 1.28 mas, which is much larger than the quoted estimated errors of either data source. We therefore conclude that the astrometric precision for most individual B-type stars in Sco-Cen is probably limited to $\sim$1 mas precision in reality, and that it cannot necessarily be stringently established at present whether the \textit{Hipparcos} or \textit{Gaia} values are more accurate.

\begin{figure}[htb]
\centering
\includegraphics[width=8cm]{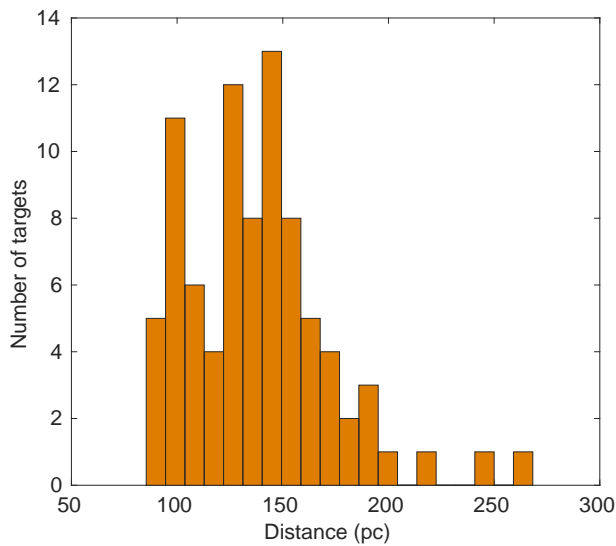}
\caption{Histogram of distances for the BEAST sample, from \textit{Gaia} where available and from \textit{Hipparcos} otherwise.}
\label{f:distances}
\end{figure}

A concrete example is the HIP 82514 system, which is an eclipsing binary. As a consequence, the radii and temperatures of the individual central stellar components can be accurately determined, leading to an unusually reliable photometric distance of $\sim$135 pc \citep{budding2015}. Meanwhile, the distance based on the \textit{Hipparcos} parallax is 154 pc, and the \textit{Gaia} DR2 distance is 268 pc. This is a clear indication that the binarity of HIP 82514 has influenced the astrometric solutions, which may have a particularly strong effect for the \textit{Gaia} analysis. Very recently, an early version of the third \textit{Gaia} data release, EDR3, has been made public. We investigated this data release in the context of the BEAST sample and found a good general consistency on the population level. EDR3 has yet smaller formal uncertainties than DR2, but this data release still only fits for parallax and proper motion, so systematic effects for binaries remain. Indeed, for HIP 82514, the EDR3 parallax-based distance is 534 pc, which is much larger than both the DR2 distance and the photometric distance. Given the large degree of multiplicity in our sample, it is not obvious that the increase in precision with EDR3 would necessarily reflect an improvement in accuracy. We thus consistently use DR2 astrometry in this study; the corresponding distances are shown in Table \ref{t:physical} and in Fig. \ref{f:distances}. Later \textit{Gaia} releases, which will attempt to account for binarity in the sources, should yield significantly better astrometric values in terms of both precision and accuracy.

Multiplicity in the sources can also affect their measured radial velocities (RVs), causing them to deviate strongly from the systemic velocity. In the standard Bayesian models such as BANYAN $\Sigma$, the model assumes that the input values correspond to the systemic values. Thus if they are substantially affected by binary motion, the membership assessment can become systematically erroneous. We attempt to account for such effects by checking how the BANYAN output is affected if an RV value is included in the analysis versus if it is not, and whether the results change significantly from the previous analysis based on \textit{Hipparcos} astrometry. While most systems show good consistency between different scenarios (as expected), there are some cases in which the outcomes are markedly different. For example, HIP 60855, HIP 65021, and HIP 76126 all had $>50$\% probabilities of membership at the selection phase, but have essentially 0\% probabilities in the updated analysis, both when RV data is included in the analysis and when it is not. Following our discussion about systematic uncertainties in the astrometric analysis, membership should not necessarily be categorically excluded in such cases; however, an individual age estimation then becomes particularly important.

The membership probabilities for each individual target are shown in Table \ref{t:astro}. The parallax and proper motion quantities and derived probabilities in the table are generally from \textit{Gaia} DR2 astrometry, except for four of the brightest targets which do not have DR2 astrometry. In these cases, \textit{Hipparcos} astrometry from \citet{vanleeuwen2007} was used instead.

\subsection{Individual ages}
\label{s:age}

The age of a target system is a crucial parameter in direct imaging studies. In particular, it is a necessary parameter in isochronal analysis for converting the detected flux from substellar companions into mass estimations, or equivalently, converting upper flux limits on non-detection cases into mass detection limits. Mass estimations and limits are in turn necessary in statistical studies for attempting to interpret the survey results in the context of theoretical predictions \citep[e.g.][]{nielsen2019,vigan2020}. However, age estimations of individual systems are often difficult and prone to large uncertainties. Individual B-type stars lack age-related chromospheric indicators and can essentially only be dated using isochronal fitting. However, B-type stars also exhibit high degrees of multiplicity and rapid rotation, both of which need to be accurately determined; otherwise ambiguities in the isochronal analysis are introduced.

In BEAST, we are greatly helped by the fact that all of the target stars (except for the outliers discussed in Sect. \ref{s:membership}) are high-probability members of Sco-Cen, which has a well-determined statistical age. The US, UCL, and LCC sub-regions have estimated ages of 10$\pm$7, 16$\pm$7, and 15$\pm$6 Myr, respectively \citep{pecaut2016}, so one option is to simply assign each stellar system the age of its associated subgroup. This would yield a broadly accurate age in most cases. However, even the sub-regions are not strictly coeval but have age differentials of a factor $\sim$2 within them \citep{pecaut2016}, so it is possible to enhance the precision for individual systems further by considering even finer-structure features within the kinematic subgroups. \citet{pecaut2016} constructed an age map of Sco-Cen based on the statistical age of the local populations of Sun-like stars, which shows continuous large-scale gradients across the Sco-Cen region. One method of more precise age determination is therefore to interpolate the age map at the location of each target. This effectively provides an age estimate that is based on isochronal dating of suitable Sco-Cen stars that are correlated particularly well with the spatial position of the target.

Another, yet more refined approach is to determine an age based on isochronal dating of Sco-Cen stars that are particularly close to the target star in phase space -- that is both in position and velocity. We executed a dedicated such procedure for the BEAST sample, which is described in the following. 

\subsubsection{Identifying co-moving stars}

As the map from \citet{pecaut2016} demonstrates, the full Sco-Cen region spans a range of ages, consistent with being the result of a complex series of smaller star formation events spanning several million years. Coeval stars within the group have correlated positions as a result, which can be utilized when extracting ages based on location in the map. However, stars arising from different formation events gradually intermix over time, so the sky-projected location of a Sco-Cen star is in principle an incomplete proxy for its age. This implies that improvements in age precision and/or accuracy could be achieved by accounting not only for correlations in physical space, but also for correlations in velocity space. 

Thus, in this section we examine each BEAST target for the purpose of trying to identify nearby stars with very similar 2D-projected space motions. We refer to such stars as co-moving stars (CMS). A CMS could be a wide, physically bound companion to the target star, but it does not need to be; we do not attempt to distinguish which CMS are physical companions and which are not. Rather, we see the identified groups of stars as co-moving streams originating from a common extended star formation event, and which will likely form the seeds for distinct young moving groups \citep[e.g.][]{zuckerman2004,torres2008} at older ages, once Sco-Cen as a whole has dispersed beyond straightforward recognition. \textit{Gaia} is of critical importance for this analysis, since it provides proper motions for a large number of Sun-like and lower-mass members of Sco-Cen that were previously impossible to identify, which can now be subjected to CMS analysis. In principle, usage of RVs would further enhance the analysis, since it would add a dimension of movement and also enable convergence tracing. However, since the vast majority (90\%) of the CMS candidates in our analysis do not have RV measurements yet, such an analysis cannot be performed at the present time. 

We start our CMS analysis by transforming the \textit{Gaia} proper motions and parallaxes into sky-projected velocities $v_{\alpha}$ and $v_{\delta}$ and heliocentric distances $d$. The physical separation $\xi$ between a candidate CMS with distance $d_i$ and a target star with distance $d_0$ can then be calculated through

\begin{equation}
\xi = \sqrt{d_i^2+d_0^2-2 d_i d_0 cos(\Delta \theta)}
,\end{equation}

where $\Delta \theta$ is the angular separation between the two objects on the sky. Likewise, we can calculate a 2D-projected differential velocity $\Delta v$ between the candidate CMS (velocity $v_i$) and target (velocity $v_0$) as

\begin{equation}
\Delta v = \sqrt{(v_{\alpha,i} - v_{\alpha,0})^2+(v_{\delta,i} - v_{\delta,0})^2} .
\end{equation}

Similarly as in, for example \citet{roser2018} and \citet{meingast2019}, we set a threshold of $\Delta v < 1.3$~km/s for distinguishing between CMS and non-CMS candidates. Since 1 km/s corresponds approximately to 1 pc/Myr and Sco-Cen is approximately 15 Myr old, it follows that only stars within $\sim$20~pc of a given target star can be a viable CMS, so we can set as an additional threshold that $\xi < 20$ pc. Finally, to avoid relying on potential CMS stars with large uncertainties, we exclude stars for which either $ \left |\frac{v_\alpha-1.3}{\sigma_{v_\alpha}}\right |<2$ or $\left | \frac{v_\delta-1.3}{\sigma_{v_\delta}} \right |<2$, that is stars that cannot be distinguished from the threshold at more than 2$\sigma$ confidence. Likewise, we exclude stars with $\frac{5}{\pi ln 10} \cdot \sigma_\pi > 0.07$, that is stars for which the uncertainty in the distance would necessitate absolute magnitude errors greater than 0.07 mag. The latter criterion is motivated by the fact that we need good precision photometry for the isochronal analysis.

\begin{figure}[htb]
\centering
\includegraphics[width=9cm]{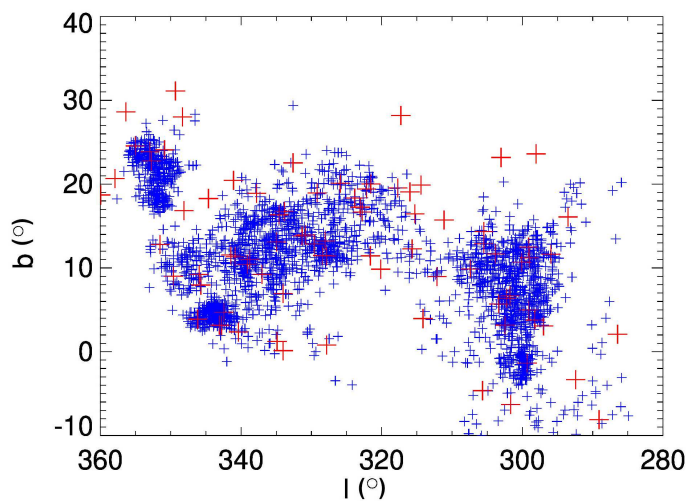}
\caption{Sky distribution of all the targets in BEAST (red symbols) and their identified CMS candidates (blue symbols).}
\label{f:scocensky}
\end{figure}

\begin{figure}[htb]
\centering
\includegraphics[width=9cm]{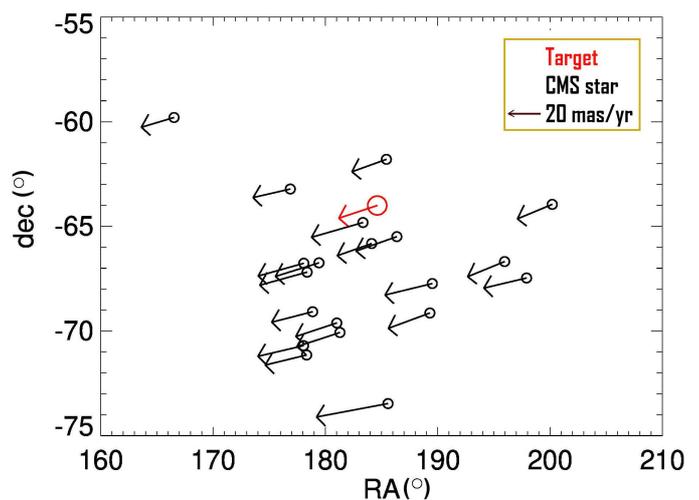}
\caption{Sky placement and proper motion vectors of HIP 60009 (in red) and its associated CMS candidates (in black). The length of the arrows is proportional to the speed of the proper motion, where the size of the arrow in the legend denotes 20 mas/yr.}
\label{f:hip60009cms}
\end{figure}

We show the distribution of targets and CMS in Fig. \ref{f:scocensky}. Incidentally, their placements also cleanly outline the three sub-regions of Sco-Cen. A specific example of a target (HIP 60009) and its associated CMS candidates is shown in Fig. \ref{f:hip60009cms}.

\subsubsection{Isochronal analysis of co-moving stars}

For each candidate CMS, we collected photometry from \textit{Gaia} \citep{brown2018} and \textit{2MASS} \citep{skrutskie2006}. We created four independent sets $\mathcal{F}$ of photometric pairs: $\{G,J\}$, $\{G,H\}$, $\{G,K\}$, and $\{Gbp,Grp\}$. We corrected for reddening using the STILISM \citep{capitanio2017} 3D maps to acquire an $E(B-V)$ value for each individual CMS, adapting the reddening to each relevant photometric band using conversions from \cite{wang2019}. Each photometric pair can be compared with theoretical isochrones in order to constrain the age and mass of a given CMS candidate. The BT-SETTL \citep{allard2014} isochrones are used for this purpose. The data points are chosen as pairs since two parameters are required to resolve the ambiguity between mass and age, and the pairs are chosen on the basis of combining a large wavelength span with small photometric errors. In the $k$-th photometric band, we can now relate the measured flux $F_k$ to model fluxes $F_{k,model}$. Given that our model can be formulated in a compact form as a 3D matrix, whose elements $\{ a_{ijk} \}$ represent the fluxes corresponding to the $i$-th mass, the $j$-th age, and the $k$-th filter, we can construct a 2D distance matrix whose elements $\delta_{ij}$ are defined as

\begin{equation}
\delta_{ij}= \sqrt{\sum_{k \in \mathcal{F}}{\left ( \frac{F_k-F_{ijk,th}}{\Delta F_k} \right )^2}}
,\end{equation}

where $\Delta F_k$ is the photometric error. In this context, $\delta_{\rm min} = \min{\delta_{ij}}$ represents the best approximation to the measured fluxes, yielding the most probable mass and age of the CMS. The upper threshold for constituting an acceptable fit was set at $\delta_{min} = 3$ as a general criterion. A CMS was also rejected in the $k$-th band if its flux error was above 10\% (0.1 mag).

With an average number of $\sim$36 suitable CMS candidates per target (see Table \ref{t:litmult}), it is in principle possible to get very precise age estimates for the targets by averaging the isochronal ages of their respective CMS samples. However, several considerations need to be taken into account for acquiring maximally accurate estimates. As a first step, we note that the CMS matching to the targets is based on a sharp cut-off of CMS versus non-CMS candidates, but in reality there is a gradient where the probability gradually decreases outward from the closest, most probable CMS companions. We account for this by calculating a weighted mean of the CMS ages, imposing weights $w_i$ defined as

\begin{equation}
w_i=\left ( \left (\frac{\Delta v_{max}}{\xi_{max}}\xi \right )^2+(\Delta v_\alpha)^2+(\Delta v_\delta)^2 \right )^{-1/2}.
\end{equation}

In this equation $\Delta v_{max} = 1.3$ km/s and $\xi_{max} = 20$ pc are the maximum allowed relative velocity and separation between a target and CMS as described in the previous section. Since the isochrones are sampled logarithmically in age, we calculate a weighted mean $m$ on the logarithmic ages rather than the linear ages (equivalent to a geometric mean) as follows:

\begin{equation}
m = \sum_i {ln(t_i) \tilde{w}_i}
,\end{equation}

where $t_i$ are the individual ages and $\tilde{w}_i$ the normalized weights (i.e. normalized to a sum of 1). The corresponding uncertainty $s$ then becomes

\begin{equation}
s = \sqrt{\sum_i \tilde{w}_i (ln(t_i) - m)^2 \cdot \frac{n}{n-1} \sum_i {\tilde{w_i}^2}}
,\end{equation}

where $\frac{n}{n-1}$ is the Bessel correction factor for CMS sample sizes of $n>1$. Since we perform the calculations on four different filter pairs, we get four different means and uncertainties $m_j$ and $s_j$. This can be translated back into linear quantities through $\mu_j = \exp{(m_j)}$ and $\sigma_j = \mu_j (\exp{(s_j)}-1)$. For a collapsed estimate of the age of the target, we combine the four cases with weights using the same approach as above, but this time using $s_j$ for the weights; $w_j = 1/s_j^2 = 1/(ln(\sigma_j / \mu_j) + 1)$. The result could in principle be taken as a final age estimate of the system, if the isochrones are sufficiently accurate. However, there are known biases in the sample that skew the isochronal analysis unless properly accounted for. 

In particular, there are two primary aspects of the target sample that can potentially impose biases in the CMS-determined ages: the age-mass bias and the multiplicity bias. The former is based on the observed fact that low-mass stars in young stellar regions such as Sco-Cen feature systematically younger isochronal age estimates than Sun-like and higher-mass stars \citep[e.g.][]{pecaut2016}. The discrepancy gets increasingly pronounced towards later spectral types across the M-type sequence. A priori, this could hypothetically be interpreted as a true mass-age gradient in the systems, with low-mass stars forming later than higher-mass stars. However, the discrepancy persists even within physically bound stellar systems that should be expected to be coeval \citep{asensio2019}. This implies that young low-mass stars have a systematic bias in their isochronal age estimates, which theoretically can be understood as an incomplete treatment in conventional models of the influence of magnetic fields on the size and temperature on the stars \citep{somers2015,feiden2016}. Since many of our CMS are low-mass stars (approximately reflecting the mass function in Sco-Cen), it is important to calibrate our procedure to account for this effect. Ideally, this would be done using models that perfectly account for the magnetic fields and the resulting chromospheric structure of low-mass stars. However, complete sets of such models do not exist yet. While there are isochronal models that do attempt to account for magnetic effects in late-type stars \citep{somers2020}, these models take the spot coverage of the star as an additional free parameter. This free parameter is relatively coarsely sampled in the isochrones, fundamentally unknown for the CMS candidates in our sample, and probably itself dependent on the age of the star. For these reasons, we opt to instead address the bias imposed by conventional isochronal models (in this case represented by BT-SETTL) in an empirical way. 

A straightforward way to avoid any mass-age bias relative to the \citet{pecaut2016} map estimates would be to use the same mass range for our CMS as was used for constructing the map (i.e. $\sim$0.8~$M_{\rm sun}$ and higher). However, imposing such a high-mass threshold would remove most of the CMS and prevent any age analysis to be performed on targets with relatively few CMS associated with them. Instead, we calculated a statistical correction factor $\beta$ based on the targets with the most numerous samples of CMS ($n > 50$). Individual $\beta$ values for such targets were calculated as the quotient between an age estimate $t_{\rm c}$ based only on $>$0.8~$M_{\rm sun}$ CMS versus an age estimate $t_0$ based on all CMS for that target. Averaged over all $n > 50$ targets, we get $\hat{\beta} = 2.20 \pm 0.03$. This factor was then uniformly applied to all (non-thresholded) age estimates in the sample, correcting for the age-mass bias.

The multiplicity bias in our sample refers to the fact that any unresolved multiplicity makes a CMS look artificially brighter, which in turn leads to a bias in its age estimation. We already discussed this as an important limitation for isochronal dating of the B-stars themselves. For the lower-mass CMS candidates, the multiplicity fraction is lower than for high-mass stars \citep{duchene2013}; therefore the effect is smaller, but still present if not accounted for. Since a typical CMS is in the M-type spectral range, we can expect a total multiplicity fraction of $\sim$30\%, and out of these multiples, we can expect $\sim$30\% to have a magnitude difference between the primary and secondary of $<$1 mag \citep{janson2012,janson2014}, at which point the secondary could start to contribute substantially to the apparent brightness of the unresolved system. In an isochronal analysis, such a system seems younger than it really is. From the multiplicity fraction and characteristic brightness differences, it follows that we can expect a small fraction of the systems to be heavily influenced by the multiplicity bias, while the majority of the sample are negligibly affected or unaffected by this effect. On this topic, it can also be noted that a small but non-negligible fraction of the CMS can be expected to host protoplanetary discs, which are more common among low-mass stars than for higher-mass stars at Sco-Cen ages \citep{carpenter2006,luhman2020}. The presence of such a disc sometimes causes a CMS to appear fainter and thus older in an isochronal analysis, producing outliers in the opposite direction from the multiplicity bias. 

Owing to the outlier nature of both of these biases, they could be easily accounted for by taking a median (rather than a mean) of the CMS ages to get a representative age estimate for the corresponding target star. However, this has the drawback that it would not be possible to use the weighted mean scheme outlined above, which was developed to increase the accuracy of the ages. Therefore, we opt for an intermediate solution based on percentile rejection: For each CMS distribution, we reject the top 10\% and bottom 10\% estimated ages (except if $n \leq 6$) and calculated a weight mean based on the remaining values to establish the best-fit age of each target star. This retains the robust properties of the median in rejecting outlier ages, whilst simultaneously keeping the weighting scheme intact. 

\begin{figure}[htb]
\centering
\includegraphics[width=9cm]{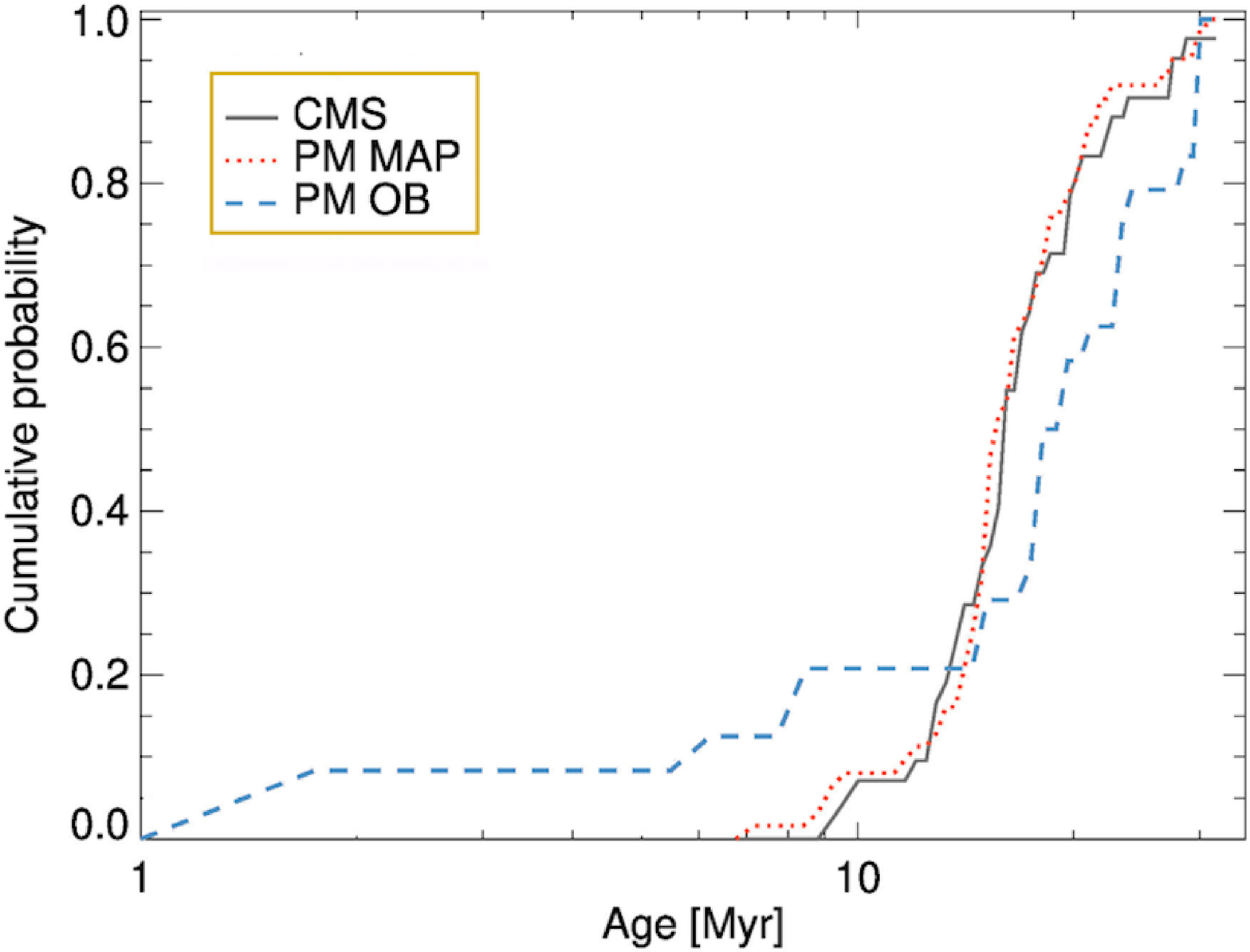}
\caption{Cumulative distributions for the target ages as estimated through our CMS analysis (black line) and through the \citet{pecaut2016} age map (red line). The two methods show a high degree of consistency. Also shown as a blue line is the distribution of ages from a direct isochronal analysis for some of the B-type stars in the sample, as determined in \citet{pecaut2016} based on models from \citet{ekstrom2012}. This latter method is less precise for a typical star in this sample relative to the other methods, and thus not used in this paper.}
\label{f:cumulative}
\end{figure}

The ages resulting from this analysis including bias mitigations are shown in Table \ref{t:ages}. The average age across the sample for this method is 16.5$\pm$0.7 Myr, which is fully consistent with average age from the \citet{pecaut2016} map of 16.2$\pm$0.6 Myr. The fact that these values are consistent is unsurprising, given that the mean age in our analysis is in some sense partially calibrated against the \citet{pecaut2016} sample in the age-mass bias correction described above. However, the fact that the cumulative distributions for the two samples show excellent consistency (see Fig. \ref{f:cumulative}) is an independent reassurance that the two methods provide statistically consistent results. Indeed, a Kolmogorov-Smirnov statistical test shows that the underlying distributions are equivalent at $>$95\% confidence. 

\subsubsection{Age summary}

In Table \ref{t:ages}, we summarize age estimations for the target sample using three different methods: (1) Average subgroup age (SG method); (2) age based on interpolation of the \citet{pecaut2016} map (MAP method); and (3) isochronal dating of co-moving stars (CMS method). The different methods complement each other rather well since, for example not all targets have identified co-moving stars associated with them, and not all stars are covered by the \citet{pecaut2016} map. In cases for which all three are available, we plan to use CMS as the first priority, MAP as second priority, and SG as third priority for statistical purposes. For individual targets with well-determined multiplicity and rotational properties, direct isochronal dating is of course another option.

\subsection{Stellar masses}
\label{s:mass}

While the precise stellar mass in a given system is not critical to determine the properties of planets in the system (unlike the system age), it is an important parameter for statistical purposes; this is the case both for the purpose of determining planet occurrence as a function of stellar mass and determining mass ratios between detected companions and their hosts. A complicating factor in determining stellar mass in a B-type sample is that the multiplicity fraction for such systems is very high; the total multiplicity is 50--60\% or higher according to the literature review of \citet{duchene2013}.
If a stellar multiple is unresolved with an unknown mass/flux ratio of the two (or more) components, then its component masses cannot be unambiguously derived from isochronal analysis. However, if a multiple system can be distinguished and monitored over a sufficient baseline, it can become possible to determine dynamical masses without any model uncertainty.

\begin{figure}[htb]
\centering
\includegraphics[width=8cm]{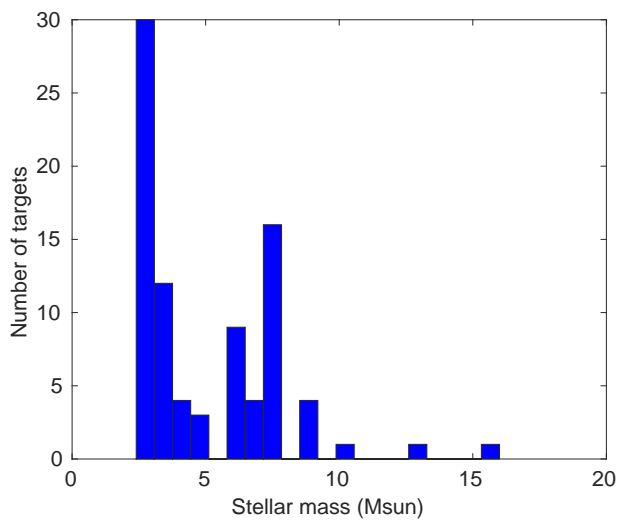}
\caption{Histogram of estimated stellar masses for the BEAST sample.}
\label{f:masses}
\end{figure}

While many of the Sco-Cen B-stars are multiple (see Sec. \ref{s:multiplicity}), most of these have no dynamical mass determinations as of yet. Hence, to estimate stellar masses in our sample, we used the spectral type (SpT) of each system as a proxy for the mass of its primary star. For this purpose, we used exactly the same relation that has already been used for early-type stars in Sco-Cen in \citet{lafreniere2014}. The resulting mass estimations are summarized in Table \ref{t:physical} and in Fig. \ref{f:masses}. 

To test the accuracy of these estimations for the BEAST sample, we cross-checked them against the three systems of the sample that are both known eclipsing binaries and double-lined spectroscopic binaries, such that individual masses can be determined in an entirely model-free manner: HIP 74950 \citep{budding2015}, HIP 78168 \citep{david2019}, and HIP 82514 \citep{budding2015}. The primary masses as determined dynamically are 4.16, 5.58, and 8.3~$M_{\rm sun}$ respectively, while the corresponding SpT-inferred primary masses are 3.3, 5.9, and 9.0 $M_{\rm sun}$. For all the tested cases, the estimated masses are thus within 6--21\% of the measured value without any noticeable systematic offsets, so we consider this to be sufficient precision for the purpose of primary mass estimation. 

We note that for most close binaries, for which any imaged planetary companion would be circumbinary, very little is known about the mass of the secondary star. This can lead to large uncertainties if the mass ratio is calculated with respect to the total central mass, as seen for HIP 79098 (AB)b in \citet{janson2019}. On a longer timescale, it is therefore highly desirable to better characterize the orbits of the close binaries in our sample. Radial velocity, \textit{Gaia} DR$>$2, interferometry, and high-resolution imaging could all be useful techniques for this purpose.

\subsection{Multiplicity}
\label{s:multiplicity}

To assess multiplicity over a wide range of semi-major axes, a combination of several different methods must be applied. For close-in binaries (periods of days to years), RV is the most applicable technique, although the broad spectral lines of B-type stars limit the achievable precision and make it more difficult to identify double-lined binaries. While there is no published coherent survey for RV multiplicity among B-stars across all of Sco-Cen, many of the targets have been observed over at least a few epochs with RV, sometimes stretching back several decades, owing to their high visual brightnesses. We scanned the literature for indications of RV binarity both in surveys and individual studies and found 22 such cases. As already mentioned in Sect. \ref{s:mass}, three of these binaries simultaneously exhibit double lines and are eclipsing; in these cases, detailed system parameters can be derived. We also identified two other eclipsing binaries in the sample from the literature search: HIP 67464 has an eclipse periodicity of 2.6 days \citep{dubath2011}, and is also listed as a spectroscopic binary in the SB9 catalogue of spectroscopic binary orbits \citep{pourbaix2004} with the same period, although seemingly only as an SB1. HIP 65112 is a system for which eclipses have been reported \citep{malkov2006,dubath2011}, but for which no reported instances of RV binarity exist in the literature. Essentially all of the close RV binaries offer the opportunity to determine unique component masses and similar constraints over short timescales in principle, but more data than is currently available would be required, for example from \textit{Gaia} astrometry or interferometric imaging. 

Intermediate-separation binaries (a few AU to hundreds of AU) are primarily detected with interferometry in the closer-in cases and AO imaging in the wider cases. In both circumstances, the WDS catalogue lists companions from the literature with a high degree of completeness. We removed known companions in the 0.1--6$^{\prime \prime}$ range from the BEAST sample as discussed in Sect. \ref{s:selection}. However, systems that only contain companions with smaller separations than 0.1$^{\prime \prime}$ were kept in the sample. These have primarily been detected through interferometry \citep{rizzuto2013}, but in some cases also with AO \citep[e.g.][]{shatsky2002}. The literature AO surveys also yielded some confirmed physical companions outside of 6$^{\prime \prime}$. Since previous surveys were not fully completed across the Sco-Cen region, we also found a number of new binary companions in our BEAST observations in the 0.1--6$^{\prime \prime}$ region where they would have been removed from the sample if they had been known beforehand. These new binaries are discussed in more detail in Sect. \ref{s:binaries}.

Wide binaries at separations in the thousands of AU can only be detected in wide-field astrometric surveys such as with \textit{Gaia}, by identifying objects that share a common proper motion (CPM) with the target star. As we have seen in Sect. \ref{s:age} however, there are many stars in Sco-Cen that share very similar proper motions with our individual target without necessarily being bound. In this paper we do not attempt to distinguish between wide binaries and other CMS stars. But in Table \ref{t:litmult}, which summarizes the (known) multiplicity properties of the sample, we also make a note of the number of CMS stars associated with each target from our isochronal analysis for general reference. The search volume for that analysis had a 20~pc radius, so most CMS stars are not expected to be physically bound.

In principle, proper motion accelerations over time for individual targets between \textit{Hipparcos} and \textit{Gaia}, for example, could also be used to identify and characterize companions \citep[e.g.][]{calissendorff2018,brandt2018,kervella2019}. However, this requires a very high precision in general, so given the large excess uncertainties discussed in Sect. \ref{s:membership}, the DR2 data are not yet sufficient for a robust analysis of the BEAST sample; subsequent \textit{Gaia} releases should be better suited for that purpose.

\subsection{Discs}
\label{s:disk}

Beyond their direct connection to planet formation and evolution, the presence and properties of discs among target stars in direct imaging studies are relevant for several purposes: The presence of bright discs may be correlated with the presence of wide massive planets \citep{meshkat2017}, and if there are signs of gaps in the discs, that may give prior information about where the planets could be located \citep[e.g.][]{apai2008,janson2013a}. Furthermore, the disc may be detected in high-contrast imaging if it is sufficiently bright, allowing for its detailed morphology to be mapped; this can provide further information about planets in the system \citep[e.g.][]{chiang2009,dong2018} and about the disc itself \citep{boccaletti2015,milli2017}.

For our sample, there are two main ways to identify discs: The dust in the disc gives rise to some degree of infrared excess, and if there is gas in the disc, it gives rise to emission lines, generally leading to a Be-type spectral classification for the star. We scanned the literature for both kinds of indicators. Infrared excess is usually inferred from \textit{Spitzer} or \textit{WISE} data, or both \citep[e.g.][]{luhman2012,rizzuto2012}. We excluded cases in which the inferred properties from the spectral energy distribution (SED) fitting are odd (e.g. unrealistically high extinction) and there is only one data point to indicate the excess. The SED fitting is relatively difficult for B-type stars, partly because the star itself is very bright even at infrared wavelengths so that a relatively large excess is required to be detectable and partly because the multiplicity rate is high, which complicates the fitting. Nonetheless, we inferred a reasonable and significant excess for 21 of the BEAST targets, which is 25\% of the sample.

Be-type stars are generally classified as such in SIMBAD, but since these classifications sometimes date back to sources for which the underlying data is not necessarily presented, we require that a star shows significant emission in at least one of the epochs presented in \citet{arcos2017} to count the star as a disc candidate in our sample. Both the excesses and Be-type classifications are shown in Table \ref{t:physical}. While line emission seems less frequent (six instances, 7\% of the sample) than excess, the two properties correlate well, in the sense that stars with identified emission almost always occur in systems with identified excess. This fact supports the notion that both properties are probably good indicators for discs in the systems. In both cases, the percentages given should be seen as merely indicative and should be regarded as lower limits, because the literature surveyed is not necessarily complete and our rejection process was conservative in order to include only reasonably secure cases.

\section{Observations}
\label{s:obs}

The BEAST survey started running during mid-2018, and has consistently used SPHERE \citep{beuzit2019} for its observations. At the time of writing, most first-epoch data (81\%) have been acquired, while almost all of the second epoch observations are allocated but remain to be executed.\footnote{The VLT has been closed down during much of 2020 as a result of the COVID-19 pandemic.} 

All first-epoch observations are acquired in the IRDIFS-EXT mode, in which the $YJH$-band wavelength range is covered by the integral field spectroscopy (IFS) arm of the instrument, and the $K$-band range is covered by the IRDIS arm. The IFS arm provides integral field spectroscopy in a $\sim 1.7 \times 1.7$ arcsec field of view (FOV) close to the target star, while IRDIS provides dual-band photometry in a larger $\sim 11 \times 11$ arcsec FOV. With its coverage in the $K$-band range through the $K1$ and $K2$ intermediate-band filters, the IRDIFS-EXT setting allows for detection of even very dusty and red planetary companions, which are a common occurrence in young regions like the Sco-Cen \citep[e.g.][]{rameau2013b,chauvin2017}. The N-ALC-YJH-S coronagraph and  pupil tracking are used during the observations to facilitate contrast enhancement in the post-processing with ADI-based techniques. All observations are carried out in service mode at the VLT, with the constraint that they have to be executed around an hour angle of zero to maximize field rotation for ADI purposes. Since the observing blocks are 75 min including overhead, this yields total field rotations in the range of a few tens of degrees for each target.

The direct integration time for each individual target was chosen to be just below the point where the star would saturate around the coronagraphic mask edges. Shorter integrations than this would compromise overheads and read noise sensitivity, while longer integrations would sacrifice sensitivity at the smallest separations near the inner working angle ($\sim$100 mas). The integration times are based on a statistical prediction of ambient conditions; during the actual observations, the conditions can be such that some saturation still occurs at the coronagraph edge. This is considered an acceptable trade-off. 

Before and after each main ADI sequence, a non-saturated image of the primary star (using neutral density filters) is acquired, and a ``waffle'' image is also acquired, in which the star is behind the mask but the waffle mode of the deformable mirror is turned on, such that ghost images appear at specific locations that can be used for both astrometric and (in principle) photometric calibration. A set of sky frames is also acquired, which is particularly important in IRDIFS-EXT observations since the IRDIS arm operates in the $K$-band range where the thermal background is relatively high. Dithering of the IRDIS detector is used to account for detector effects such as bad pixels.

Since Sco-Cen as a whole is relatively close to the Galactic plane, background sources are commonplace, therefore follow-up observations need to be executed in the majority of cases to distinguish background stars from real physical companions. Most follow-up observations will be executed in an identical way as the first-epoch observations, with occasional exceptions such as if a disc candidate is detected, in which case the follow-up may be performed at a shorter wavelength and with polarimetric differential imaging \citep[PDI; see e.g.][]{hashimoto2011,schmid2018} in addition to ADI. Since a primary goal with the second epoch is generally to test for  CPM, and since Sco-Cen members typically have a relatively modest proper motion of 20--30 mas/yr, we aim for all second epoch observations to be acquired with at least a one-year baseline w.r.t. the first epoch. Further follow-up, such as detailed spectroscopic characterization, can then be done in a third epoch once CPM is established;  these follow ups are usually outside of the large programme, in which the main part of
the programme is run. If the companion happens to reside in the IFS FOV, then spectroscopic characterization is possible even in the first and second epochs of data. In this paper we focus primarily on the first epoch observations.

An observing log of the first epoch observations is shown in Table \ref{t:log}. In ESO's service mode, the observations are associated with a set of constraints on the ambient conditions that need to be fulfilled for the output data to be scientifically useful. Observations  are typically only executed if all the constraints are fulfilled; if conditions deteriorate significantly during an observation, it is discarded and re-executed at a later time. In our case, the most central constraint is that atmospheric conditions should have a grade of ``good'' or better in the ESO classification scheme, which corresponds to a seeing better than 0.9$^{\prime \prime}$ and a coherence time better than 4 ms\footnote{For the earliest observations, the seeing criterion was 0.8$^{\prime \prime}$, but we relaxed it slightly from 2019 onwards to allow for easier scheduling.}, apart from a timing constraint requiring that the observations must take place around a meridian transit. In the ESO grading system, ``A'' means that all the constraints were fulfilled, while ``B'' means only small deviations from the requirements. Observations graded ``C'' or lower are considered not scientifically useful and scheduled for re-execution. In total, 67 first-epoch observations with an ``A'' or ``B'' grading have been acquired to date, which are included with their respective grading in the observing log. 

\section{Data reduction and analysis}
\label{s:data}

The fundamental data reduction process for the survey is performed in a primarily automatic and streamlined manner, using the SPHERE Data Center \citep[DC; see][]{delorme2017}. Basic data reduction steps include dark and flat corrections; pixel scale, true north corrections, and distortion corrections from observations of stellar clusters; image cube creation; and wavelength calibration. The true north angle is typically -1.8 deg and is determined on a run-by-run basis with a precision of $\sim$0.1 deg or better. Similarly, the pixel scale is typically 12.25 mas/pixel for IRDIS and 7.46 mas/pixel for IFS and these quantities are determined each run with a precision of 0.01--0.02 mas/pixel. After the calibrated image cubes are produced, the high-contrast processing begins, which uses the SpeCal package \citep{galicher2018} within the context of DC as well. SpeCal allows for a wide range of possible options in terms of optimization algorithms and corresponding parameter settings. 

As part of our standard reduction procedure, we use three separate algorithms for IRDIS and three for IFS for each target. For IRDIS, the three approaches are as follows: (1) A rotation and collapse of the time series without any ADI subtraction, which allows for a very quick overview of the field and the data quality. (2) A classical ADI approach with a median across the time series representing the point spread function (PSF) model, which yields a conservative reduction that has benefits when searching for faint extended emission (debris discs) that can easily get partially subtracted in a more aggressive scheme. (3) A Template Localized Combination of Images (TLOCI) approach, which is optimized for finding point sources in the images. The IFS procedure is similar, but because of its normally entirely contrast-limited FOV, a non-ADI approach is not very meaningful. Thus the three approaches are as follows: (1) a classical ADI approach; (2) a TLOCI approach; and (3) a Karhunen-Lo{\`e}ve Image Projection \citep[KLIP, see][]{soummer2012,pueyo2016} Principal Component Analysis (PCA)-based approach as an alternative to the TLOCI algorithm. Beyond this standardized procedure, several efforts are in progress to analyse the BEAST data with alternative reduction schemes, including PynPoint \citep{stolker2019}, ANDROMEDA \citep{cantalloube2015}, and TRAP \citep{samland2020}, but this paper focusses on the standardized procedure.

After the final reduced images are produced, they are visually inspected and convincing point sources at $\geq$5$\sigma$ are identified. All of these companion candidates (CCs) are analysed with the characterization module of SpeCal, in a second step of DC processing. This yields $K1$ and $K2$ photometry (the two bands in the IRDIS dual-band data), calibrated astrometry for all CCs in the IRDIS FOV, and extracted spectra for all CCs that are close enough to the star to also be included in the IFS FOV; this is a vast minority of all CCs. The $K1-K2$ photometry can be used to identify, for example particularly red candidates, but like most individual dual-band colours, this photometry normally does not produce very distinctive results, since there is a substantial overlap between the expected colours of planets and the colour distribution of background stars. The only way to definitively confirm (or reject) a given CC is to follow it up in a second epoch, taking both the proper motion and the colour information into account simultaneously.

\section{Results and discussion}
\label{s:results}

\subsection{Contrast}
\label{s:contrast}

\begin{figure}[htb]
\centering
\includegraphics[width=8cm]{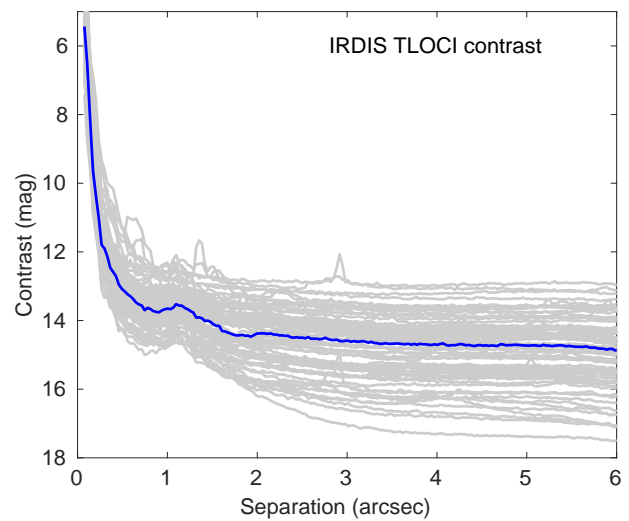}
\caption{Contrast curves for IRDIS with the TLOCI reduction. The $K1$ images alone are used for this purpose owing to the limited sensitivity of the $K2$ images. Grey lines: Contrast curves for individual targets. Blue thick line: Median contrast curve for the sample.}
\label{f:irdiscontrast}
\end{figure}

\begin{figure}[htb]
\centering
\includegraphics[width=8cm]{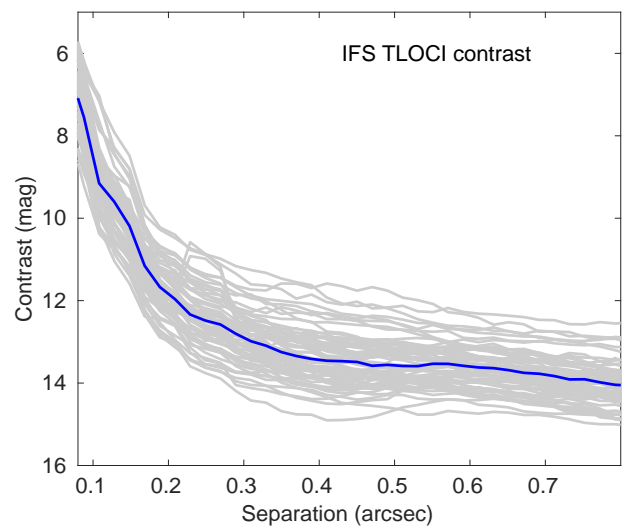}
\caption{Contrast curves for IFS with the TLOCI reduction. The collapsed images across the accessible wavelength range is used. Grey lines: Contrast curves for individual targets. Blue thick line: Median contrast curve for the sample.}
\label{f:ifscontrast}
\end{figure}

We show the 5$\sigma$ contrast performance as a function of separation for IRDIS in Fig. \ref{f:irdiscontrast} and for IFS in Fig. \ref{f:ifscontrast}. For example, at 0.4$^{\prime \prime}$, the median contrast across the observed sample is $9 \times 10^{-6}$ (12.6 mag) for IRDIS and $4 \times 10^{-6}$ (13.5 mag) for IFS. This is deeper thanthe GPIES \citep{nielsen2019} and SHINE \citep{langlois2020} surveys, for instance. The contrast performance results from a combination of the excellent performance of SPHERE, the relatively good (but not exceptional) average observing conditions, and the high average brightness of the stars in the sample, allowing for a good wavefront sensing performance in each case. This contrast resulted from a baseline SpeCal reduction optimized for uniformity across the sample. It would most likely be possible to enhance the contrast further still on a star-by-star basis through alternative high-contrast algorithms \citep[e.g.][]{samland2020} or through adjusting the parameters in the SpeCal algorithm. As mentioned in Sect. \ref{s:data}, experiments to this end are ongoing in parallel to the survey-scale reduction efforts. The BEAST sample is nearly ideal for such experiments owing to the widely extended contrast-limited regimes of the bright host stars.

\subsection{Faint companion candidates}
\label{s:ccs}

\begin{figure}[htb]
\centering
\includegraphics[width=8cm]{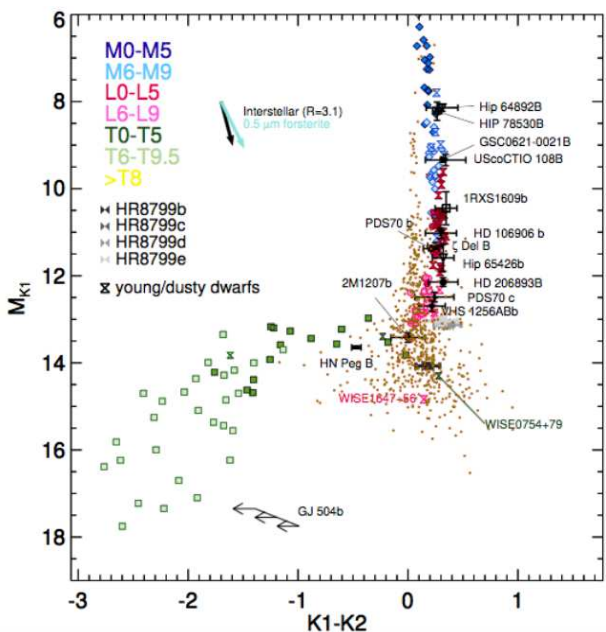}
\caption{Colour-magnitude diagram in $M_{K1}$ vs. $K1-K2$ for all candidates in the BEAST first-epoch observations performed to date. The CCs are denoted with brown dots, while template objects representing various spectral types are indicated with different colours as shown in the legend. Previously discovered low-mass substellar companions that have been observed in the $K1$ and $K2$ filters are also shown as black and grey symbols. For relatively bright CCs, background stars can be distinguished from probable physical companions in most cases, but for fainter CCs, there is substantial overlap between the two populations in the diagram.}
\label{f:cmd}
\end{figure}

The observations presented in this work contain a total of 708 faint candidates inside the IRDIS FOV of the 67 observed targets. As expected from the varying stellar backgrounds, these candidates are unevenly distributed among the targets, of which 11 have empty fields without any candidates; the most crowded field has 117 candidates. All of the candidates will require follow-up to test for CPM in a second epoch before it can be categorically established whether they are physical companions or background stars. This was expected from the outset and is a part of the allocated BEAST observational programme. Since the dual-band imaging mode of IRDIS is used, $K1$ and $K2$ images exist for all fields and could in principle be used to distinguish physical companions from background stars in a colour-magnitude diagram (CMD), even from a single epoch. However, as we noted in Sect. \ref{s:data}, the $K1-K2$ colour is usually not particularly distinctive. Some of the faintest candidates also lack a $K2$ detection, since the background noise is higher in $K2$ than in $K1$. Nonetheless, particularly high-merit candidates may potentially be identified through unusually blue (implying molecular absorption) or red (implying clouds) colours. A collective CMD for all candidates with measured $K1-K2$ colours is shown in Fig. \ref{f:cmd}.

\subsection{Binary companions in the SPHERE images}
\label{s:binaries}

When a point source around a target star is bright enough to be a stellar companion, it has an extremely low false positive probability in general, so it can already be regarded as a probable companion from the first epoch. We thus present the astrometric properties of the companions that have been detected in the stellar mass/brightness regime in Table \ref{t:newbins}\footnote{With the exception of HIP 76600, see individual note.}. Their statistical properties will be examined in more detail in Squicciarini et al. (in prep). In total, we identified ten stellar companions, of which six had not previously been identified in the scientific literature. We briefly discuss each individual case below.

\begin{itemize}

\item \textbf{HIP 50847:} The presence of a stellar companion to HIP 50847 was tentatively hinted at in an observation taken for a multiplicity study of Sco-Cen presented in \citet{janson2013b}. However, the corresponding data set was taken at ambient conditions far inferior to the programme requirements, and therefore the reduced image was of so poor quality that no actual inference could be made about the veracity of such a companion. The BEAST data set is the first in which a companion can be robustly inferred (see Fig. \ref{f:hip50847system}). Aside from the visual binarity, HIP 50847 is also known as a double-lined spectroscopic binary with a period of only 15 days \citep{quiroga2010}, meaning that there are three confirmed stellar components in the system. 

\begin{figure}[htb]
\centering
\includegraphics[width=8cm]{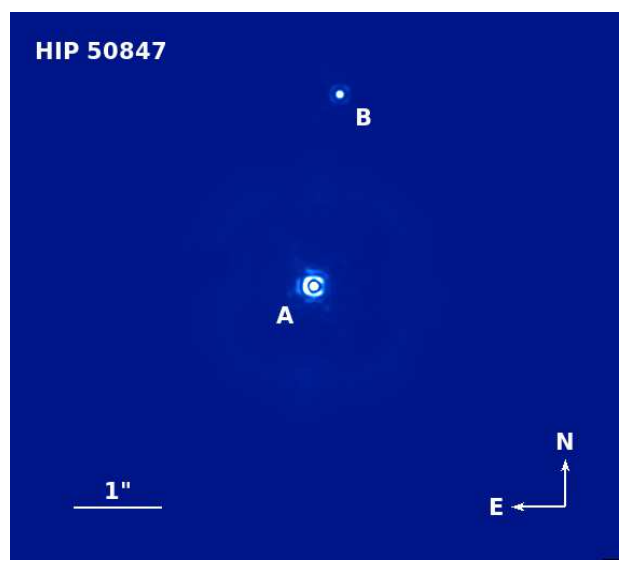}
\caption{Non-ADI IRDIS image of HIP 50847, with the primary component denoted ``A'' and the new companion at intermediate separation denoted ``B''.}
\label{f:hip50847system}
\end{figure}

\item \textbf{HIP 59173:} This star was previously observed with AO imaging in a survey by \citet{shatsky2002}. The companion detected in BEAST is not reported in \citet{shatsky2002}, which is probably because the companion is a low-mass star. Its contrast relative to the primary star may have been challenging for the first-generation AO system ADONIS used in that survey. In the literature, HIP 59173 is noted as a double-lined spectroscopic binary \citep{chini2012}; no period is reported, but since the lines can evidently be kinematically distinguished in the RV data, the spectroscopic binarity must refer to a much closer-in companion than the BEAST-detected companion at $\sim$171 AU projected separation, and so the system has a triple or higher-order multiplicity.

\item \textbf{HIP 60009:} Very similarly to HIP 59173, HIP 60009 was observed with ADONIS in the \citet{shatsky2002} survey, but the companion detected in BEAST was not reported in their study. In this case, presumably the small separation of $\sim$150 mas of the companion would have made it difficult to detect with ADONIS. Also very similarly to HIP 59173, HIP 60009 has been noted as a double-lined spectroscopic binary in the literature \citep{chini2012}, so as a whole the system is probably at least triple. An image of the stellar system is shown in Fig. \ref{f:hip60009system}.

\begin{figure}[htb]
\centering
\includegraphics[width=8cm]{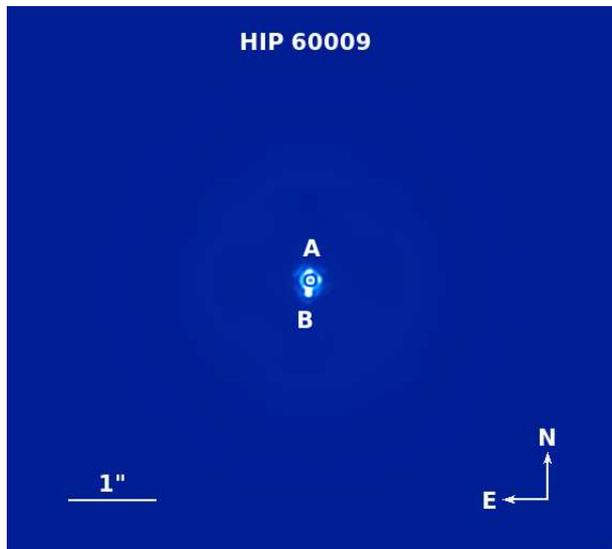}
\caption{Non-ADI IRDIS image of HIP 60009, showing the primary (``A'') component as well as the newly discovered close companion (``B'').}
\label{f:hip60009system}
\end{figure}

\item \textbf{HIP 62434:} According to our archival search, HIP 62434 has never been observed with AO or similar facilities, and thus its intermediate-separation multiplicity has been largely unprobed. As we note in Table \ref{t:litmult}, this object has been subject to RV studies in the past \citep{pourbaix2004}, which have revealed a spectroscopic companion with a period of 1828 days. Our detected companion from BEAST images has a projected separation of $\sim$357 AU and is therefore certainly distinct from the RV companion. Hence, the HIP 62434 system is at least a stellar triple.

\item \textbf{HIP 63005:} Similar to HIP 59173 and HIP 60009, HIP 63005 was observed in the \citet{shatsky2002} AO survey, but the companion detected in the BEAST data was not reported there. The BEAST companion resides at 274 mas separation, which would have made it difficult to see with ADONIS. HIP 63005 is in a very wide binary pair with the B-star HIP 63003, which is also part of the survey; otherwise there are no additional stellar companions known around either star.

\item \textbf{HIP 63945:} The companion discovered in BEAST was previously known, as was reported in \cite{shatsky2002}. Normally, such an identified intermediate-separation stellar binary would have been discarded in the BEAST master list construction, but in this case it had been overlooked owing to the relatively complex multiplicity listing for this particular target in the WDS database. One of the reasons for this is that the system additionally has an inner interferometrically detected companion \citep{rizzuto2013}, which was close-in enough to be allowed within the BEAST selection criteria. HIP 63945 also has a noted single-line spectroscopic binary companion of undetermined period \citep{chini2012}, which may or may not be identical to the interferometric companion. Like many of the other visual binaries reported in this paper, the system is thus a high-order multiple with at least three components.

\item \textbf{HIP 73624:} HIP 73624 was originally included in the BEAST sample, but when a bright stellar companion was discovered in the IRDIS FOV during a short aborted run, it was deemed a suboptimal target and removed from the sample. Hence, this target is not listed in other BEAST tables, but for completeness we include the location of the newfound stellar companion in Table \ref{t:newbins}.

\item \textbf{HIP 74100:} This target has a companion noted in \citet{shatsky2002}, but it was kept in the original sample because the quoted separation in the WDS catalogue was well within the range of where the companion would have impacted the observational performance of SPHERE. Since a companion was detected at a 550 mas separation in BEAST, we revisited the literature. We note that the separation, 6 mas, quoted in \citet{shatsky2002} must presumably be a typographical error because the observations were performed with ADONIS whose angular resolution was far worse than $\sim$6 mas. It seems probable that the companion detected in BEAST is the same as that in \citet{shatsky2002} for HIP 74100, but we assess that the literature astrometry cannot be relied on for, for example CPM or orbital analysis for this target. 

\item \textbf{HIP 76600:} There is an interferometric companion to this target noted in \citet{rizzuto2013}, which is marginally detected in BEAST images, however not to the extent that reliable astrometric properties can be determined. We therefore omit this target from table \ref{t:newbins}. There is also a spectroscopic companion with a period of only 3.3 days \citep{pourbaix2004}, which is therefore a distinct third stellar component in the system.

\begin{figure}[htb]
\centering
\includegraphics[width=8cm]{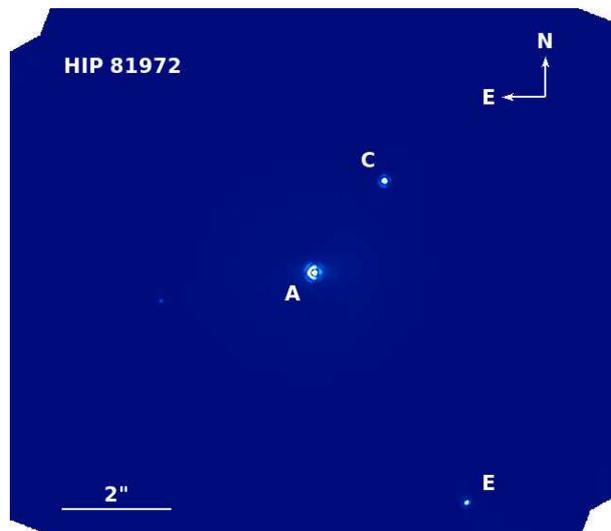}
\caption{Non-ADI IRDIS image of HIP 81972. The component notation follows the notation from previous epoch imaging of the system \citep{shatsky2002}. We find that ``C'' is a background star, while ``E'' is a real physical companion.}
\label{f:hip81972noadi}
\end{figure}

\item \textbf{HIP 81792:} HIP 81792 resides in an unusually crowded region of the sky, and thus has a much higher spatial density of bright contaminants relative to most other targets. In the IRDIS FOV, there are two candidates that are both bright enough to be stellar even if co-moving; HIP 81972 is the only target in the observed sample for which this is the case. Because of these special circumstances, we cannot necessarily infer that they are probable companions despite their high brightness. Fortunately, HIP 81972 was observed by \citet{shatsky2002} and both candidates are reported as visible in their data, so we have the opportunity to test whether they are physical companions through a CPM test. We designate the candidates ``C'' and ``E',' respectively, following the notation in WDS. The CPM test for ``C'' is shown in Fig. \ref{f:hip81972bg}, and the test for ``E'' in Fig. \ref{f:hip81972cpm}. Somewhat counter-intuitively, the data show that the brighter and closer-in candidate ``C'' is most likely a background contaminant, while the slightly fainter and more distant HIP 81972 E appears to be a genuine physical companion to HIP 81972 A. Being identified as a likely background source, ``C'' is not included in Table \ref{t:newbins}; its separation and position angle in the 2019 BEAST epoch are 2107.1$\pm$7.2 mas and 322.92$\pm$0.23 deg, respectively. Since the epoch 2000 data presented in \citet{shatsky2002} were taken with ADONIS and pre-date the ESO data archive, it cannot be easily validated, so it is desirable to double-check the CPM and non-CPM conclusions against the second-epoch BEAST data, which is also the plan for all the other candidates. As a side note, the ``B'' component in WDS for this system (which is outside of the IRDIS FOV) is denoted as being possibly physical in the WDS notes. But the ``B'' component is sufficiently separated from HIP 81972 A to have its distinct \textit{Gaia} entry (Gaia DR2 5968351361342059264) and its proper motion and parallax reported in \textit{Gaia} imply that it is a physically unrelated background source. We show an image of the HIP 81972 system in Fig. \ref{f:hip81972noadi}.

\end{itemize}

\begin{figure}[htb]
\centering
\includegraphics[width=8cm]{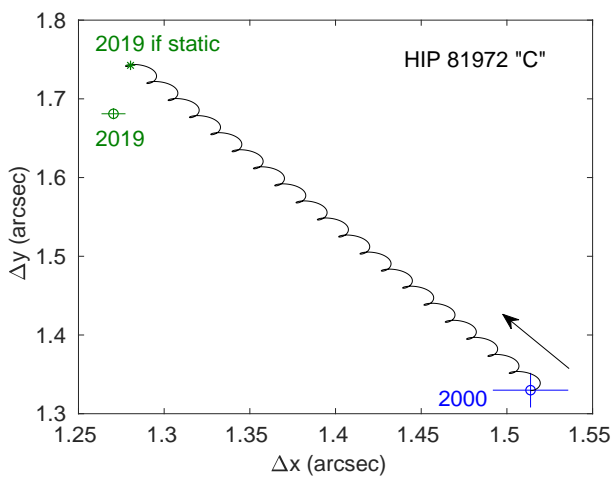}
\caption{Proper motion test for the ``C'' candidate around HIP 81972. The black arrow denotes the direction of time for background motion. The motion of the candidate from the 2000 to the 2019 epoch is inconsistent with CPM, but can be explained by a background star with a small proper motion vector of its own.}
\label{f:hip81972bg}
\end{figure}

\begin{figure}[htb]
\centering
\includegraphics[width=8cm]{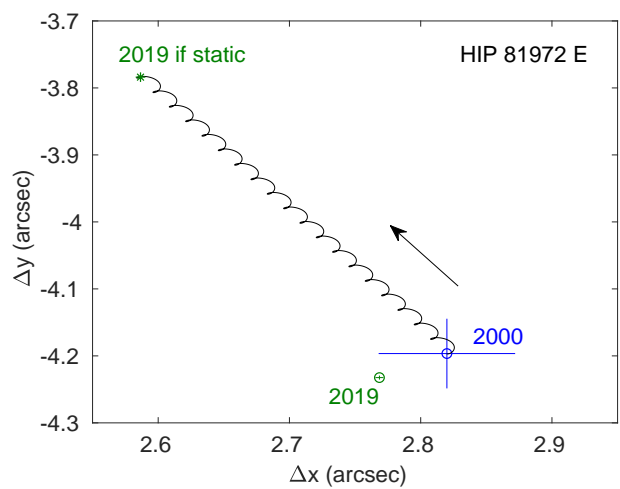}
\caption{Proper motion test for the ``E'' candidate around HIP 81972. The black arrow denotes the direction of time for background motion. For this case, the candidate is consistent with CPM, while simultaneously being clearly distinct from the static background location. It is a probable physical companion.}
\label{f:hip81972cpm}
\end{figure}

\section{Conclusions}
\label{s:summary}

The first large-scale, extreme-AO survey of planetary systems around B-type stars, BEAST will probe a nearly complete sample of suitable B-type stars in the Sco-Cen region. This paper explores and summarizes the properties of the BEAST stellar sample. Important parameters for both interpreting individual systems and statistical interpretations on a population basis include the stellar masses, multiplicity properties, disc properties, and ages. We place particular emphasis on the age aspect, where the Sco-Cen region offers some particularly appealing opportunities for determining accurate ages. We develop a scheme for age determination of the B-type stars in the sample based on isochronal dating of stars that can be closely kinematically matched to each individual target. This allows for a high precision in the age estimations, since an average of several co-moving stars can be acquired. The resulting ages are statistically consistent with other age estimations from the literature, but as expected they are generally more precise and most likely more accurate.

First epoch observations were performed for 67 out of the 85 targets in the survey, and these observations yielded several hundreds of candidate companions, which will need to be followed up in a second epoch to test for CPM. As in all direct imaging surveys, the majority of the candidates are expected to be physically unrelated background stars. However, if the frequency of wide giant planets continues to increase with stellar mass in the same way as the frequency increases in the M- to A-type stellar range, then BEAST could potentially yield more new planet discoveries than any other direct imaging survey ever performed. The contrast limits provided in the survey are deeper on average than in comparable surveys, owing in large part to the performance of the SPHERE instrument, in particular for the bright stars that constitute the targets in the survey.

We presented six new and four previously known binary companions detected within the survey. Since these detections are very bright, their false alarm probability is typically very low, owing to the low probability of having a sufficiently bright star in the background within a few arcseconds of the target. An exception is the HIP 81972 system, which resides in front of an unusually crowded field, and as a result, does have a bright background object in the IRDIS field of view. Previously reported in the literature as HIP 81972 C, we  tested the object for CPM using literature astrometry to create a 19-year baseline and found that it is consistent with a near-static background star. On the other hand, another star in the field, HIP 81972 E, is fully consistent with being physically bound to the B-star primary.

\begin{acknowledgements}
M.J. gratefully acknowledges funding from the Knut and Alice Wallenberg Foundation. This study made use of the CDS services SIMBAD and VizieR, and the SAO/NASA ADS service. R.G. and S.D. are supported by the project PRIN-INAF 2016 The Cradle of Life - GENESIS-SKA (General Conditions in Early Planetary Systems for the rise of life with SKA). We also acknowledge support from INAF/Frontiera (Fostering high ResolutiON Technology and Innovation for Exoplanets and Research in Astrophysics) through the ``Progetti Premiali'' funding scheme of the Italian Ministry of Education, University, and Research. Part of this research was carried out at the Jet Propulsion Laboratory, California Institute of Technology, under a contract with the National Aeronautics and Space Administration (NASA). E.E.M. acknowledges support from the Jet Propulsion Laboratory Exoplanetary Science Initiative and the NASA NExSS Program. This work has made use of the SPHERE Data Centre, jointly operated by OSUG/IPAG (Grenoble), PYTHEAS/LAM/CeSAM (Marseille), OCA/Lagrange (Nice), Observatoire de Paris/LESIA (Paris), and Observatoire de Lyon/CRAL, and is supported by a grant from Labex OSUG@2020 (Investissements d’avenir – ANR10 LABX56). G-DM acknowledges the support of the DFG priority program SPP 1992 ``Exploring the Diversity of Extrasolar Planets’’ (KU 2849/7-1) and from the Swiss National Science Foundation under grant BSSGI0$\_$155816 ``PlanetsInTime''. Parts of this work have been carried out within the framework of the NCCR PlanetS supported by the Swiss National Science Foundation. P. Delorme is supported by the French National Research Agency in the framework of the Investissements d'Avenir program (ANR-15-IDEX-02), through the funding of the ``Origin of Life'' project of the Univ. Grenoble-Alpes.
\end{acknowledgements}

\begin{table}[htb]
\tiny
\caption{Observing log for BEAST first epoch data up through 2020.}
\label{t:log}
\centering
\begin{tabular}{lllrlr}
\hline
\hline
Target & RA & Dec & Date & Grade$^a$ & Rot.$^b$ \\
 & (hh mm ss) & (dd mm ss) & & (A/B) & (deg) \\
\hline
HIP 50847       &       10 22 58.15     &       -66 54 05.4     &       27 Jan 2019        &       B       &       17.6    \\
HIP 52742       &       10 46 57.47     &       -56 45 25.9     &       25 Jan 2019        &       B       &       22.1    \\
HIP 54767       &       11 12 45.21     &       -64 10 11.2     &       11 Mar 2019        &       A       &       18.5    \\
HIP 58452       &       11 59 10.68     &       -45 49 56.0     &       8 Apr 2018        &       A       &       31.9    \\
HIP 58901       &       12 04 45.25     &       -59 15 11.8     &       19 Feb 2019        &       B       &       20.3    \\
HIP 59173       &       12 08 05.22     &       -50 39 40.6     &       11 Mar 2019        &       A       &       26.1    \\
\hline
\end{tabular}
\begin{list}{}{}
\item[$^{\mathrm{a}}$] ESO rating of ambient conditions, see text for details.
\item[$^{\mathrm{b}}$] Field rotation between first and last useful exposure (pupil stabilized mode).
\end{list}
\end{table}

\clearpage
\onecolumn

\begin{longtable}{llllrrrrlr}
\caption{Subgroup membership of the BEAST sample.}\\
\hline\hline
Target & Alt name & RA & Dec & Plx & $\sigma_{\rm plx}$ & $\mu_{\rm RA}$ & $\mu_{\rm Dec}$ & Region & Prob. \\
 & & (hh mm ss) & (dd mm ss) & (mas) & (mas) & (mas/yr) & (mas/yr) & & (\%) \\
\hline
\endfirsthead
\caption{continued.}\\
\hline\hline
Target & Alt name & RA & Dec & Plx & $\sigma_{\rm plx}$ & $\mu_{\rm RA}$ & $\mu_{\rm Dec}$ & Region & Prob. \\
 & & (hh mm ss) & (dd mm ss) & (mas) & (mas) & (mas/yr) & (mas/yr) & & (\%) \\
 \hline
\endhead
\hline
\endfoot
\hline
\multicolumn{10}{l}{\scriptsize{$^{\mathrm{a}}$Unreliable value for HIP 82514; see Sect. \ref{s:membership}}}\\
\hline
\endlastfoot

HIP 50847       &       L Car   &       10 22 58.15     &       -66 54 05.4     &       8.16    &       0.20    &       -22.23  &       12.17   &       LCC     &       79.9    \\
HIP 52742       &       HR 4221 &       10 46 57.47     &       -56 45 25.9     &       4.98    &       0.23    &       -20.71  &       0.57    &       LCC     &       0.0     \\
HIP 54767       &       HD 97583        &       11 12 45.21     &       -64 10 11.2 &       10.59   &       0.14    &       -47.32  &       5.46    &       LCC     &       3.9     \\
HIP 58452       &       HD 104080       &       11 59 10.68     &       -45 49 56.0 &       7.52    &       0.07    &       -30.43  &       -9.03   &       LCC     &       46.8    \\
HIP 58901       &       HD 104900       &       12 04 45.25     &       -59 15 11.8 &       9.14    &       0.08    &       -38.22  &       -12.04  &       LCC     &       99.3    \\
HIP 59173       &       HR 4618 &       12 08 05.22     &       -50 39 40.6     &       9.85    &       0.32    &       -34.35  &       -11.51  &       LCC     &       99.1    \\
HIP 59747       &       $\delta$ Cru    &       12 15 08.72     &       -58 44 56.1 &       11.62   &       0.96    &       -37.29  &       -9.97   &       LCC     &       99.6    \\
HIP 60009       &       $\zeta$ Cru     &       12 18 26.25     &       -64 00 11.1 &       10.49   &       0.33    &       -34.96  &       -6.24   &       LCC     &       99.9    \\
\end{longtable}


\begin{longtable}{lrrr}
\caption{Age estimations for the sample.}\\
\hline\hline
Target & CMS age$^{\rm a}$ & MAP age$^{\rm b}$ & SG age$^{\rm c}$ \\
 & (Myr) & (Myr) & (Myr) \\
 \hline
\endfirsthead
\caption{continued.}\\
\hline\hline
Target & CMS age$^{\rm a}$ & MAP age$^{\rm b}$ & SG age$^{\rm c}$ \\
 & (Myr) & (Myr) & (Myr) \\
 \hline
\endhead
\hline
\endfoot
\hline
\multicolumn{4}{l}{\scriptsize{$^{\rm a}$ Age based on co-moving stars; see text.}}\\
\multicolumn{4}{l}{\scriptsize{$^{\rm b}$ Age based on the map in \citet{pecaut2016}.}}\\
\multicolumn{4}{l}{\scriptsize{$^{\rm c}$ Age of the closest matching Sco-Cen subgroup.}}\\
\hline
\endlastfoot
HIP 50847       &       27.3$\pm$1.3    &       26.7    &       15$\pm$6        \\
HIP 52742       &       82.5$\pm$21.8   &       \textemdash     &       15$\pm$6        \\
HIP 54767       &       84.5$\pm$8.1    &       22.4    &       15$\pm$6        \\
HIP 58452       &       20.0$\pm$0.6    &       \textemdash     &       15$\pm$6        \\
HIP 58901       &       13.9$\pm$0.5    &       17.1    &       15$\pm$6        \\
HIP 59173       &       13.7$\pm$0.6    &       20.6    &       15$\pm$6        \\
HIP 59747       &       \textemdash     &       15.9    &       15$\pm$6        \\
HIP 60009       &       19.8$\pm$0.9    &       16.2    &       15$\pm$6        \\
\end{longtable}


\begin{longtable}{lllllr}
\caption{Summary of known companions and co-moving stars}\\
\hline\hline
Target & SB/EB$^{\rm a}$ & Reference & AO/INT$^{\rm b}$ & Reference & CMS$^{\rm c}$ \\
\hline
\endfirsthead
\caption{continued.}\\
\hline\hline
Target & SB/EB$^{\rm a}$ & Reference & AO/INT$^{\rm b}$ & Reference & CMS$^{\rm c}$ \\
 \hline
\endhead
\hline
\endfoot
\hline
\multicolumn{6}{l}{\scriptsize{$^a$Spectroscopic (SB) and eclipsing (EB) binary companions.}}\\
\multicolumn{6}{l}{\scriptsize{$^b$Companions identified with  AO or interferometry (INT).}}\\
\multicolumn{6}{l}{\scriptsize{$^c$Number of identified candidate co-moving stars. Most are not physically bound.}}\\
\multicolumn{6}{l}{\scriptsize{$^d$Interferometric companion from \citet{rizzuto2013} and AO companion from \citet{shatsky2002}.}}\\
\hline
\endlastfoot
HIP 50847       &       SB2     &       \citet{quiroga2010}     &       AO      &       This work    &       18      \\
HIP 52742       &       \textemdash     &       \textemdash     &       \textemdash     &       \textemdash     &       1       \\
HIP 54767       &       \textemdash     &       \textemdash     &       \textemdash     &       \textemdash     &       4       \\
HIP 58452       &       \textemdash     &       \textemdash     &       \textemdash     &       \textemdash     &       21      \\
HIP 58901       &       \textemdash     &       \textemdash     &       \textemdash     &       \textemdash     &       172     \\
HIP 59173       &       SB2     &       \citet{chini2012}       &       AO      &       This work    &       52      \\
HIP 59747       &       SB2     &       \citet{chini2012}       &       \textemdash     &       \textemdash     &       0       \\
HIP 60009       &       SB2     &       \citet{chini2012}       &       AO      &       This work    &       14      \\
\end{longtable}


\begin{longtable}{lrlrrrll}
\caption{Fundamental properties of the BEAST sample.}\\
\hline\hline
Target & Dist & SpT & Mass & App. V & App. K & Disc? & Ref.$^{\rm a}$ \\
 & (pc) &  & $M_{\rm sun}$ & (mag) & (mag) & &  \\
 \hline
\endfirsthead
\caption{continued.}\\
\hline\hline
Target & Dist & SpT & Mass & App. V & App. K & Disc? & Ref.$^{\rm a}$ \\
 & (pc) &  & $M_{\rm sun}$ & (mag) & (mag) & &  \\
 \hline
\endhead
\hline
\endfoot
\hline
\multicolumn{8}{l}{\scriptsize{$^{\rm a}$ References. A17: \citet{arcos2017}. C09: \citet{carpenter2009}. C16: \citet{cotten2016}.}}\\
\multicolumn{8}{l}{\scriptsize{L12: \citet{luhman2012}. L14: \citet{liu2014}. R12: \citet{rizzuto2012}.}}\\
\multicolumn{8}{l}{\scriptsize{$^{\rm b}$ Unreliable value for HIP 82514; see Sect. \ref{s:membership}}}\\

\hline
\endlastfoot
HIP 50847       &       122.6   &       B8V     &       2.9     &       4.99    &       5.314   &               &               \\
HIP 52742       &       200.8   &       B8.5III &       2.7     &       5.23    &       5.225   &       Be-type &       A17     \\
HIP 54767       &       94.5    &       B8V     &       2.9     &       5.22    &       5.418   &               &               \\
HIP 58452       &       133.1   &       B8/B9V  &       2.7     &       6.34    &       6.529   &               &               \\
HIP 58901       &       109.4   &       B9V     &       2.5     &       6.20    &       6.278   &               &               \\
HIP 59173       &       101.5   &       B6III   &       3.7     &       4.47    &       4.871   &       Excess  &       R12     \\
HIP 59747       &       86.0    &       B2IV    &       7.8     &       2.78    &       3.532   &               &               \\
HIP 60009       &       95.3    &       B2.5V   &       6.8     &       4.05    &       4.535   &               &               \\
\end{longtable}


\begin{table}[htb]
\tiny
\caption{Properties of binary companions in the BEAST data.}
\label{t:newbins}
\centering
\begin{tabular}{lllrrrrl}
\hline
\hline
Target & Pair & Date & $\rho$ (mas) & $\sigma_\rho$ (mas) & $\theta$ (deg) & $\sigma_\theta$ (deg) & Disc.$^{\rm a}$ \\
\hline
HIP 50847       &       AB      &       27 Jan 2019     &       2215.1  &       3.0     &       352.45  &       0.30    &       TW      \\
HIP 59173       &       AB      &       11 Mar 2019     &       1275.2  &       2.8     &       129.88  &       0.09    &       TW      \\
HIP 60009       &       AB      &       22 Feb 2019     &       148.2   &       0.7     &       170.89  &       0.11    &       TW      \\
HIP 62434       &       AB      &       31 Mar 2019     &       4176.1  &       3.1     &       121.84  &       0.05    &       TW      \\
HIP 63005       &       AB      &       4 Mar 2019      &       274.0   &       1.1     &       172.36  &       0.12    &       TW      \\
HIP 63945       &       AC$^{\rm b}$    &       22 Apr 2018     &       1554.0  &       6.1     &       261.61  &       0.46    &       S02     \\
HIP 73624$^{\rm c}$     &       AB      &       26 Apr 2019     &       339.3   &       1.3     &       140.70  &       0.15    &       TW      \\
HIP 74100       &       AB      &       23 Mar 2019     &       550.3   &       1.9     &       277.44  &       0.42    &       S02     \\
HIP 81972       &       AE$^{\rm b}$    &       7 Aug 2019      &       5057.5  &       3.3     &       213.19  &       0.05    &       S02     \\
\hline
\end{tabular}
\begin{list}{}{}
\item[$^{\mathrm{a}}$] Discovery reference. TW: This work. S02: \citet{shatsky2002}.
\item[$^{\mathrm{b}}$] Following component designation in WDS for previously known companions.
\item[$^{\mathrm{c}}$] The target HIP 73624 was removed from the BEAST sample during the survey; see text.
\end{list}
\end{table}

\end{document}